\documentclass[preprint,aps,superscriptaddress,floatfix,prl]{revtex4-1}
\usepackage{subfigure}
\usepackage{amssymb, amsmath}
\usepackage{graphics}

\usepackage{tikz}
\usepackage{tikzscale}
\usepackage{pgfplots}
\pgfplotsset{compat=newest}
\pgfplotsset{plot coordinates/math parser=false}
\usetikzlibrary{plotmarks}

\newcommand{\Ri}[0]{\textnormal{Ri}}
\renewcommand{\d}[0]{\textnormal{d}}
\newcommand{\fl}[1]{{#1}'}
\newcommand{\urms}[1]{u_{\textnormal{rms}}}

\newcommand{\unit}[1]{\ensuremath{\, \mathrm{\hspace{0.5mm}#1}}}

\newcommand{\figLabel}{Fig.~}

\renewcommand{\d}{\textnormal d}

\newcommand{\avx}[1]{\left\langle {#1} \right\rangle} 

\graphicspath{{figures/}}

\begin{document}

\title{Experimental study of the initial growth of a localized turbulent patch in a stably stratified fluid}
\date{\today}

 \author{Lilly \surname{Verso}}
 \affiliation{School of Mechanical Engineering, Tel Aviv University, Tel Aviv, Israel}
\email{alexlib@tau.ac.il}

 \author{Maarten \surname{van Reeuwijk}}
 \affiliation{Department of Civil and Environmental Engineering, Imperial College, London, Great Britain}

 \author{Roi \surname{Gurka}}
 \affiliation{School of Coastal and Marine Systems Science, Coastal Carolina University, Conway, SC, USA}

 \author{Peter J. \surname{Diamessis}}
 \affiliation{School of Civil and Environmental Engineering, Cornell University, Ithaca, NY, USA}

\author{Zachary J. \surname{Taylor}}
\affiliation{School of Mechanical Engineering, Tel Aviv University, Tel Aviv, Israel}

 \author{Alex \surname{Liberzon}}
 \affiliation{School of Mechanical Engineering, Tel Aviv University, Tel Aviv, Israel}

\begin{abstract}

We present a laboratory experiment of the growth of a turbulent patch in a stably stratified fluid, due to a localized source of turbulence, generated by an oscillating grid. Synchronized and overlapping particle image velocimetry and planar laser induced fluorescence measurements have been conducted capturing the evolution of the patch through its initial growth until it reached a maximum size, followed by its collapse. The simultaneous measurements of density and velocity fields allow for a direct quantification of the degree of mixing within the patch, the propagation speed of the turbulent/non-turbulent interface and its thickness. The velocity measurements indicate significant non-equilibrium effects inside the patch which are not consistent with the classical used grid-action model. A local analysis of the turbulent/non-turbulent interface provides direct measurements of the entrainment velocity $w_e$ as compared to the local vertical velocity and turbulent intensity at the proximity of the interface. It is found that the entrainment rate $E$ is constrained in the range of $0 \div 0.1$ and that the local, gradient Richardson number at the interface is $\mathcal{O} (100)$. Finally, we show that the mean flow is responsible for the patch collapse.\end{abstract}

\maketitle

\section{Introduction}\label{sec:intro}

In the stably stratified thermocline of the open ocean, turbulence is important for diapycnal transport and mixing of energy, mass, heat, nutrients and chemicals. Turbulence in the ocean drives vertical transport and mixing against the stabilizing effect of the ambient stratification through a variety of processes occurring in mid-water \cite{AlfordGregg2001, AlfordPinkel2000, Broutman1986, LiYamazaki2001, Sanford1995}. Furthermore, in regions of strong bathymetric variations \cite{Polzin1998, KlymakMoum2007, Kelly2007}, topographically generated turbulence can extend to $\mathcal{O}$(1km) above the bottom and can drive significant mixing of the strongly stratified ambient fluid \cite{GarrettLaurent2002}. These turbulent processes are evident as strong localized perturbations in horizontal and vertical small scales flow measurements \cite{Marmorino1987, Nasmyth1970}, presumably formed by the breaking of internal waves. Such intermittent signatures motivated studies of stratified \emph{turbulent patches} -- a localized volumes of turbulent activity -- as one of the prominent oceanic processes~\cite{Garrett1972, Garrett1979}. Understanding of the internal structure and mechanisms of a turbulent patch may provide insight and lead to improved parameterizations of oceanic turbulent transport and mixing \cite{Arneborg2002, Caldwell2001}.

Different types of laboratory experiments have been conducted to study turbulent patches in stably stratified fluid layers. Growth of localized turbulent sources was studied in the context of wakes of self-propelled bodies in stratified environment~\cite{vandeWatering1966, Merritt1974, Lin1979}. The studies used impulsively or continuously forced patches due to self-propelled bodies, and the main focus of these studies is on the later times of the momentumless wake evolution. Van de Watering~\cite{vandeWatering1966} mechanically created a patch in a quiescent linearly stratified flow, studying the growth rate and the upper vertical size limit of a patch. Wu~\cite{Wu1969} measured the evolution of an intrusion formed by the injection of fluid with uniform density into a stably stratified environment, which is in the vertical direction is somewhat similar to the growth of a patch. 

Fernando and co-authors \cite{Fernando1988, DeSilva1998} studied the growth of turbulent patches using a horizontally oriented oscillating grid spanning the full extent of the tank. It was found that the patch vertical size growth rate was initially unaffected by the stratification, until $Nt \approx 4$ when it started to slow down. Here $N^2 = (-g/\rho_0)d\rho/dz$ is the square of the Brunt-V\"ais\"al\"a buoyancy frequency of the ambient fluid, $g$ is the acceleration due to gravity, $\rho$ is the density and $\rho_0$ is a reference density. De Silva and Fernando~\cite{DeSilva1998} carried out laboratory experiments of a localized turbulent patch focusing on its collapse and the subsequent intrusive gravity current. Similarly to the experiments with the oscillating grid spanning the entire tank, the localized patch grew vertically until $N t \approx 4$, after which a reduction of the vertical patch extent was observed. The gravity current formed as a result of the mixing inside the patch, which created hydrostatic pressures that are larger inside the patch than in the ambient, thereby providing a forcing for the gravity current. 

The goal of this work is to better understand the early patch dynamics preceding the formation of an intrusive gravity current. We extend the previous studies by considering a patch that is free to grow in all directions. We utilize particle image velocimetry (PIV) and planar laser induced fluorescence (PLIF) techniques that enable simultaneous measurements of velocity and density fields inside and outside the patch during its initial growth phase. The data is used to provide insight into the fluid dynamics inside the patch, the buoyancy distribution and conduct a local analysis of the turbulent-non turbulent interface.

The paper is organized as follows. Section~\ref{sec:setup} presents the experimental setup and measurement techniques, specifically the details of the PIV and PLIF calibration. Section~\ref{sec:results} demonstrates the results in terms of the initial growth of the patch and its spatially averaged characteristics as well as the local analysis near the turbulent/non-turbulent interface (TNTI), aiming at processes responsible for the TNTI propagation.  Discussion and concluding remarks appear in Section~\ref{sec:conclusions}.

\section{Experimental setup} \label{sec:setup}
 
\begin{figure}
\centering
\subfigure[]{\includegraphics[width=.49\textwidth]{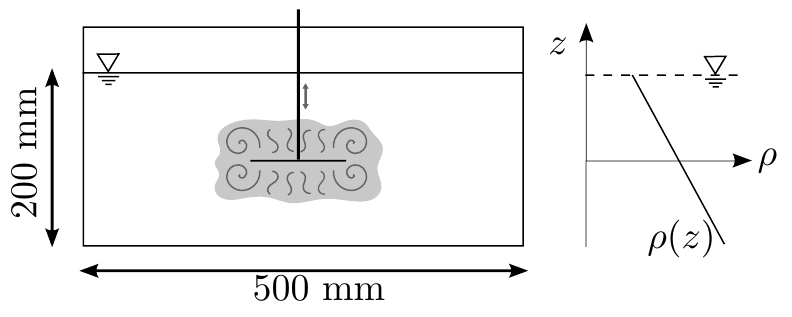}}
\subfigure[]{\includegraphics[width=.49\textwidth]{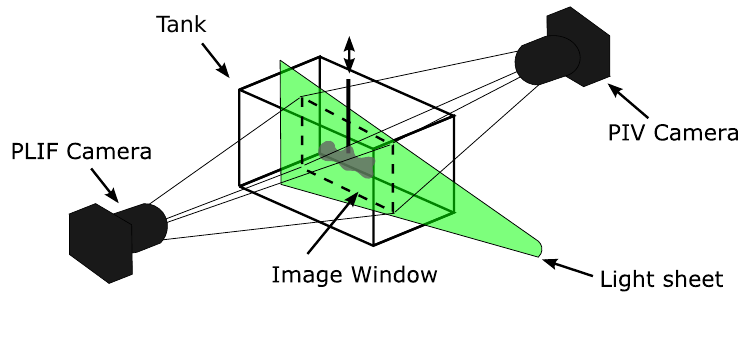}}

\caption{(a) Illustration of the size of the tank, the density gradient, and the motion of the turbulence-generating grid. (b) Illustration demonstrating the position of the PIV and PLIF cameras and the light sheet orientation with respect to the turbulent patch.}
\label{fig:exp_schematic}
\end{figure}

The problem studied here is a localized mixed turbulent region in linear stable density profile $\rho(z)$ shown schematically in \figLabel\ref{fig:exp_schematic}a. In our experimental case it corresponds to a buoyancy frequency of $N=1$ s$^{-1}$,  measured by PLIF as well as by a pycnometer with small volumes of liquid extracted at different depths. 

In order to measure the key parameters of the problem, we create a localized source of turbulence in a stably stratified environment, and measure the turbulent flow as it evolves in time. We follow the patch using optical measurement methods, starting from rest ($t=0$) through the growth phase up to its collapse. During the experiments, we maintain the refractive index matching in order to allow for accurate density and velocity measurements at high spatial resolution as described in Section~\ref{sec:index_of_refraction}.


The experiments are performed in a glass tank with a $200 \times 500$ mm$^{2}$ cross-section and a depth of 200 mm. Experiments are carried out in different solutions of an index-of-refraction matched stratified mixture of sugar, water and Epsom salts following the method of McDougall \cite{McDougall1979} described in details below. A stable linear density gradient is established in the tank using the free-flow two-tank method \cite{Economidou2009}.

%

\subsection{Turbulent agitation}\label{sec:turb_agitation}

The patch is created using an oscillating grid located at the center of the tank. The grid size is $60 \times 60$ mm$^2$ constructed of plastic square bars of $2 \unit{mm}$ and a mesh size of $10 \unit{mm}$ giving a solidity of 30.6\%. The oscillations are provided via a slide on a linear ball-bearing rail connected to a motor with an eccentric linkage. The stroke length (top of stroke to bottom of stroke) is $10 \unit{mm}$. Over the first two seconds of the oscillations, the motor is provided a ramp input to minimize any impulse from the start-up procedure and followed by a steady current input while the oscillations last for 8 seconds in total. The forcing is stopped after 8 seconds, in order to maintain the stratified layers for at least 6 sequential runs. The oscillation frequency obtained after the initial ramp input ranges between 3 to 8 Hz in increments of 1 Hz.

The patch initially grows as a result of the agitation until it reaches a maximum. The vertical patch size will be denoted $\delta$ and the typical turbulent velocity scale associated with the turbulence inside the patch will be denoted $\sigma_{u}$. Both $\delta$ as $\sigma_u$ will be defined rigorously in Section~\ref{sec:results}. In Table \ref{tab:experiment}, the measurements of the maximum patch size, $\delta_{\mathrm{max}}$ and the corresponding level of turbulence, $\sigma_{u} (\delta_{\mathrm{max}})$ are presented for the 6 different frequencies.

The key dimensionless parameters characterizing the patch dynamics are the bulk Reynolds number $Re=\sigma_u \delta / \nu$, representative of turbulence strength relative to viscosity, and the bulk Richardson number $Ri=N^2 \delta^2 / \sigma_u^2$, representative of the strength of the stratification relative to turbulence levels inside the patch. In Table \ref{tab:experiment} we report that the $Re$ ranges from 50 to 400 whilst $Ri$ ranges from 50 to 150, indicating that the stratification is relatively strong in this setup as compared to the turbulence levels.  In the results and discussion sections we will refer to the experiments by their forcing frequency, $f = 3 \div 8$ Hz, which is directly proportional to the bulk Reynolds number.

\subsection{Refractive index matching}\label{sec:index_of_refraction}

Obtaining a linear stable stratification using variations in salt concentration is a commonly used experimental technique~\cite{Fernando1988, Spedding1997, DeSilva1998}. However, when using optical methods there are some concerns about the distortion of the light path of the laser~\cite{McDougall1979} in addition to the deviation of the light paths towards the camera~\cite{Alahyari1994}.

The solution in the present case follows the work of McDougall~\cite{McDougall1979} and makes use of two liquids:
a sugar-water solution and an Epsom salt (MgS0$_4$) - water solution. The refractive index, $n$ and density, $\rho$ (at 20 $^{o}$C, measured in g/mL) can be expressed by a Taylor series following McDougall~\cite{McDougall1979} as: 
\begin{eqnarray}
n(z) &=& 1.3330 + 0.20286 C_{A}(z)+ 0.14445 C_{B}(z)+ \mathcal{O}(z^2) \\ \nonumber
\rho(z) &=& 0.9982 + 1.0054 C_{A}(z)+0.38407 C_{B}(z)+ \mathcal{O}(z^2)  
\end{eqnarray}

\noindent where $C_{A}$ and $C_{B}$ are the concentrations (weight of solute per unit weight of solution) of Epsom-salt and sugar, respectively. The desired density difference of $20 \unit{g/L}$ between the bottom of the tank and the free surface is obtained with a maximum difference in the index of refraction which compares favorably with previous studies~\cite{Daviero2001}. The maximum difference in the index of refraction of $\Delta n < 0.00001$ is obtained in each solution separately and for the two solutions being mixed in agreement with the results of McDougall~\cite{McDougall1979}.

\begin{table}[t]
\caption{Experimental parameters of different experimental runs for increasing oscillation frequency of the grid.\label{tab:experiment}}
\centering
\begin{tabular*}{0.8\textwidth}{@{\extracolsep{\fill} } c|r r r r r }
$f$ [Hz] & $\delta_{max}$ [m]  & $\sigma_{u}(\delta_{max})$ [m/s]  &  $Re$  & $Ri$  \\
\hline
\small{3} & \small{$0.026$} &\small{$2.1 \times 10^{-3}$} & \small{$55$}  & \small{$146$}  \tabularnewline 
\small{4} & \small{$0.034$} &\small{$3.6 \times 10^{-3}$} & \small{$122$}  & \small{$87$}  \tabularnewline 
\small{5} & \small{$0.038$} &\small{$4.6 \times 10^{-3}$} & \small{$179$} & \small{$69$}  \tabularnewline
\small{6} & \small{$0.043$} &\small{$5.2 \times 10^{-3}$}   & \small{$224$} & \small{$68$}  \tabularnewline 
\small{7} & \small{$0.048$} &\small{$6.3 \times 10^{-3}$} & \small{$308$} & \small{$57$}   \tabularnewline
\small{8} & \small{$0.053$} &\small{$7.4 \times 10^{-3}$} & \small{$396$} & \small{$50$}   \tabularnewline
\end{tabular*}
\end{table}

\subsection{Particle Image Velocimetry (PIV)}\label{sec:PIV}

Two-dimensional, two-component PIV measurements are performed in a vertical plane passing through the horizontal center of the tank as shown in \figLabel\ref{fig:exp_schematic}b. The PIV system in this study is composed of a Nd:YAG laser (120 mJ/pulse, wavelength of 532 nm) and a 11 MP double-exposure CCD camera with a 12-bit sensor. The imaging set-up yields a physical magnification of $56.2\,\mu$m/pixel.

The PIV image pairs are processed using the open source software described in Taylor et al.~\cite{Taylor2010} (\texttt{\small www.openpiv.net}) with interrogation windows of $32 \times 32 \unit{pixels^2}$ and 50\% overlap. Spurious vectors are identified first by a global filter and followed by a local median filter. The rejected vectors are replaced by interpolation using the local mean and the data are smoothed using a Gaussian filter.

PIV measurements are performed such that tracers (polyamid spheres, mean diameter of $50 \unit{\mu m}$, $\rho_p = 1.03$ g/cm$^3$, Dantec Inc.) have an average displacement of approximately 5 pixels for each experiment. The timing is set from time interval $\Delta t = 0.018 \unit{s}$ for the $f = 3 \unit{Hz}$ experiment down to $\Delta t = 0.006  \unit{s}$ for the $f = 8 \unit{Hz}$ run.  External triggering system initiates the motion of the grid and the first PIV image pair at the same time. In addition, the PIV measurements are synchronized with the overlapping PLIF measurements to obtain simultaneous measurements of density and velocity.

\subsection{Planar laser induced fluorescence (PLIF)}\label{sec:plif}

PLIF camera takes images simultaneously with the PIV measurements using the setup shown in \figLabel\ref{fig:exp_schematic}b. The PLIF method enables the measurement of the spatial density distribution. Rhodamine 6G is added to the mixing tank of the heavier liquid (the Epsom salt solution) at a concentration of $50 \unit{\mu g/L}$. The density can be obtained through measurements of the Rhodamine 6G concentration. This technique has previously been used in the study of interfacial mixing in stratified environments \cite{Atsavapranee1997}, 
wave breaking \cite{Troy2005}, jets \cite{Sarathi2011}, and in stratified gravity currents \cite{Odier2009}. The common assumption that the Rhodamine 6G follows the density field relies on the notion that the diffusivity of turbulence ($\kappa_T\approx \sigma_u \times l=1 \unit{cm^2 s^{-1}}$) is significantly greater than the molecular diffusivity of the Rhodamine 6G ($\kappa_{R6G}=0.
12 \times 10^{-5} \unit{cm^2 s^{-1}}$), Epsom salt ($\kappa_{MgS0_4}=0.61 \times 10^{-5}\unit{cm^2 s^{-1}}$), and sucrose ($\kappa_{s}=0.45 \times 10^{-5}\unit{cm^2 s^{-1}}$). 
Thus, the Batchelor scale $\lambda_B$ ($= \eta/Sc^{1/2}$), of the Epsom salt, which we estimate using the Kolmogorov length scale ($\eta \approx 200 \unit{\mu m}$ based on the turbulent velocity and length scales) and the Schmidt number of the Epsom salt, $Sc$ ($ = \nu/\kappa$), is found to be of the order of $10 \unit{\mu m}$.  This scale is much smaller than the size covered by one pixel ($56 \unit {\mu m}$). The under-resolution of the density field is due to the need to accommodate the large field-of-view of the experiment, following the motion of the growing patch. 

\begin{figure}
\includegraphics[width=\textwidth]{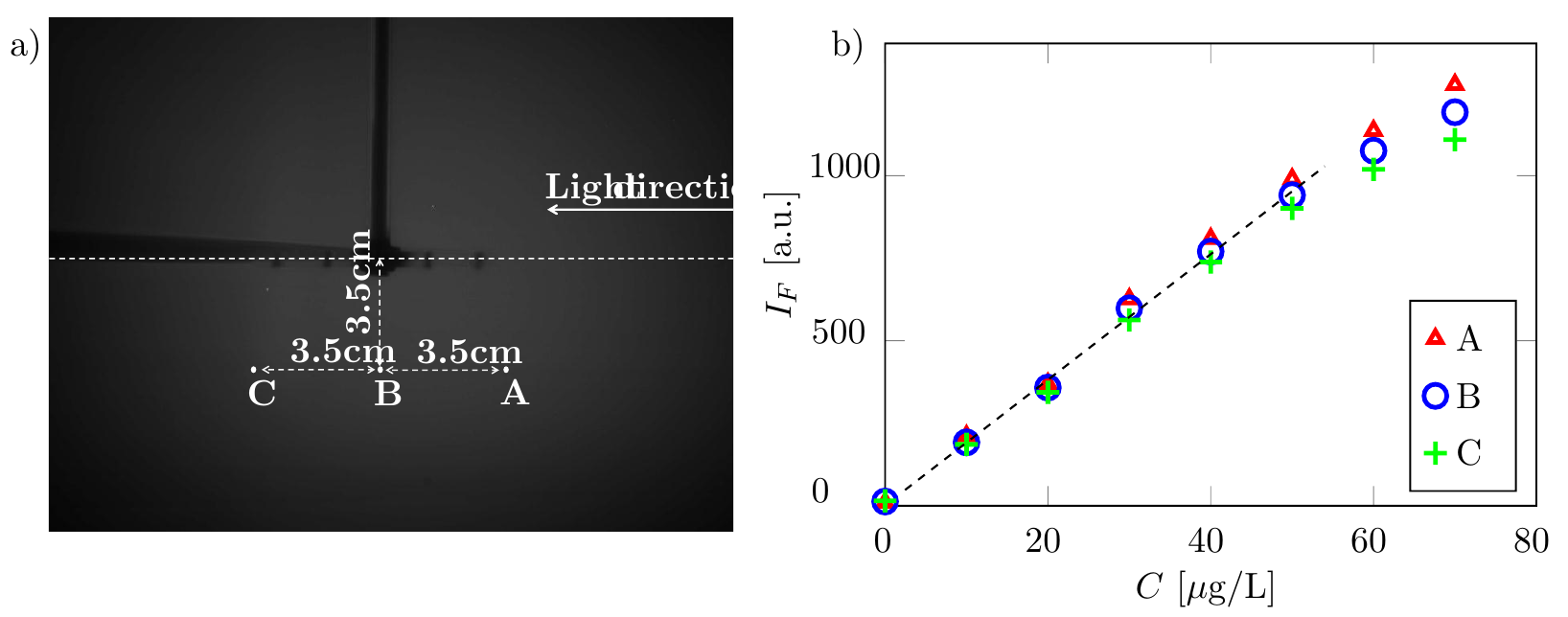}
\caption{ (a) Raw PLIF image with the laser light entering the field of view from the right. The three control points A,B, and C are used for the calibration procedure. (b) Calibration of concentration $C$ to PLIF image intensity $I_F$ obtained at the points A,B and C. Note that only the values in the linear range $C=10 \div 50 \mu$g/L are used for the calibration.} 
\label{fig:plifcal}
\end{figure}

The  CCD camera for the PLIF measurements is identical to the PIV camera (11 MP, 12 bit camera). Both cameras are equipped with 60 mm Nikkor lenses. A bandpass optical filter (Newport 20BPF10-550) with an admittance band of $550 \pm 10$ nm is fitted to the PLIF camera to transfer the light at 555 nm that Rhodamine 6G emits when excited by the Nd:YAG laser and blocks the laser light at 532 nm diffracted by the PIV tracers. 

The PLIF calibration procedure follows the method of Crimaldi~\cite{Crimaldi2008}. The intensity of light fluoresced by the dye substance $I_F$ is directly proportional to its  concentration $C$ and the intensity of light excitation, $I$, and it is inversely proportional to the ratio of the laser light intensity and the saturated intensity of the particular dye, $I_{s}$:
\begin{equation}
I_F \propto C  \frac{I}{1+I/I_{s}} 
\end{equation}
For carefully constructed experiment the intensity in PLIF images $I \ll I_s $  and consequently the Eq.~\ref{eq:IF} can be formulated as $I_F \propto C \; I$ \cite{Crimaldi2008}. Thus, the fluorescent light captured on the imaging sensor at pixel $i,j$ is a function of the pixel location (due to imperfections of imaging optics), local dye concentration $C(i,j)$ and the light intensity $I(i,j)$:
\begin{equation}\label{eq:IF}
 I_{F}(i,j)= \alpha(i,j) \; C(i,j)\;  I(i,j)
\end{equation}
where $\alpha(i,j)$ is a point-wise calibration coefficient. 

The calibration procedure goal is to obtain the calibration coefficient $\alpha(i,j)$. We use uniform solutions of seven different concentrations of Rhodamine 6G with a homogeneous mixture of Epsom salts, sucrose and water to replicate the experimental conditions. The pixel-by-pixel, concentration-independent, calibration coefficient, $\alpha(i,j)$ is obtained using an image processing procedure that is based on: i) a background image obtained as an average of the five different uniform calibration concentration fields, $I(i,j)$ ii) the dark response of the camera obtained using the images with the lens cap covered, $B(i,j)$ and iii) the step-wise correction scheme of the attenuation coefficient, $a(i,j)$ that integrates attenuation along the light rays originating from the laser source: 
\begin{equation}\label{eq:alpha}
\alpha (i,j) = \frac{1}{N_c}\sum_{c}\frac{I\left(i,j\right) - B\left(i,j\right)}{a(i,j) C}.
\end{equation} 
The laser light sheet is aligned to pass the tank horizontally (isopycnal) from one side to another. The most significant correction is calculated in a step-wise manner from the uniform concentration calibration images, from one side of the image, to the other following the light beam paths of the laser sheet, modeled using a radial coordinate system with the origin at the laser source (at the back focal position of the cylindrical lens). The radially and azimuthally varying attenuation coefficient (upon transformation of $r,\theta$ to $i,j$ for each pixel) is computed assuming a constant absorption coefficient $\beta$ of Rhodamine 6G, obtained from the work of Ferrier \cite{Ferrier1993}, $\beta = 1.1 \times 10^5$ (cm M)$^{-1}$. Constant absorption coefficient assumption was shown to be valid for concentrations below 50 $\mu$g/L:
\begin{equation}	
a(i,j) = \exp\left[-\beta\int_{r_0}^r C\left(r^{\prime},\theta\right)dr^{\prime}\right] \label{eq:theta}
\end{equation}
%
%

An example raw PLIF calibration image of the uniform concentration fluid illuminated by the laser sheet is shown in \figLabel\ref{fig:plifcal}(a). The values at three different points (A,B, and C), used for the calibration procedure example according to Eqs.~\ref{eq:IF} and \ref{eq:alpha} are shown in \figLabel\ref{fig:plifcal}(b). Note that we performed seven calibration runs with monotonically increased concentrations, but used only five that refer to the linear part of the fit, in the range of $C=10 \div 50 \mu$g/L. The variation of $ \alpha(i,j) $ across the image is smaller than 2\% for this range, as shown in \figLabel\ref{fig:plifcal}(b). 
\begin{figure}
\includegraphics[width=.9\textwidth]{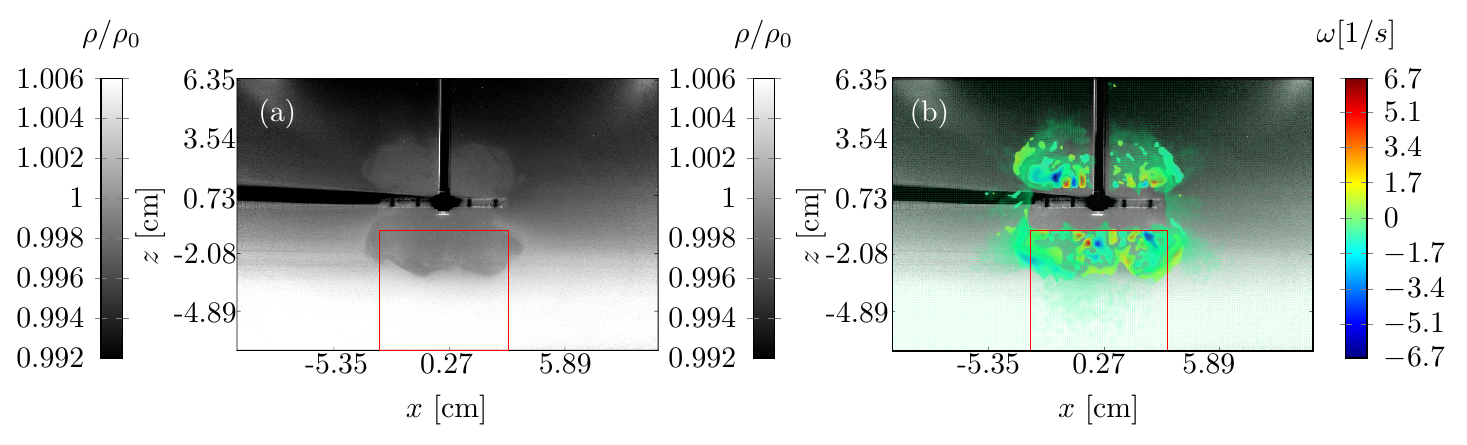}
\caption{An example of the simultaneous and overlapping PIV and PLIF measurements. The colour map corresponds to the density values, normalised to the average density $\rho_{0}$. The rectangle denotes the region of interest of the following quantitative analysis.}
\label{fig:comb}
\end{figure}

The combination of PIV and PLIF experiments allow for simultaneous measurements of density and velocity at high spatial resolution. An example of these overlapped data is shown in \figLabel\ref{fig:comb}. In this figure we also indicate the region of horizontally homogeneous part of the patch which is used later for the quantitative spatially averaged measurements of the patch size, interface thickness, turbulence kinetic energy, vorticity and the buoyancy distributions. 


\section{Results}\label{sec:results}

\subsection{Evolution of the patch interface}\label{sec:patch_evolution}

\begin{figure}
\centering
\includegraphics[width=\textwidth]{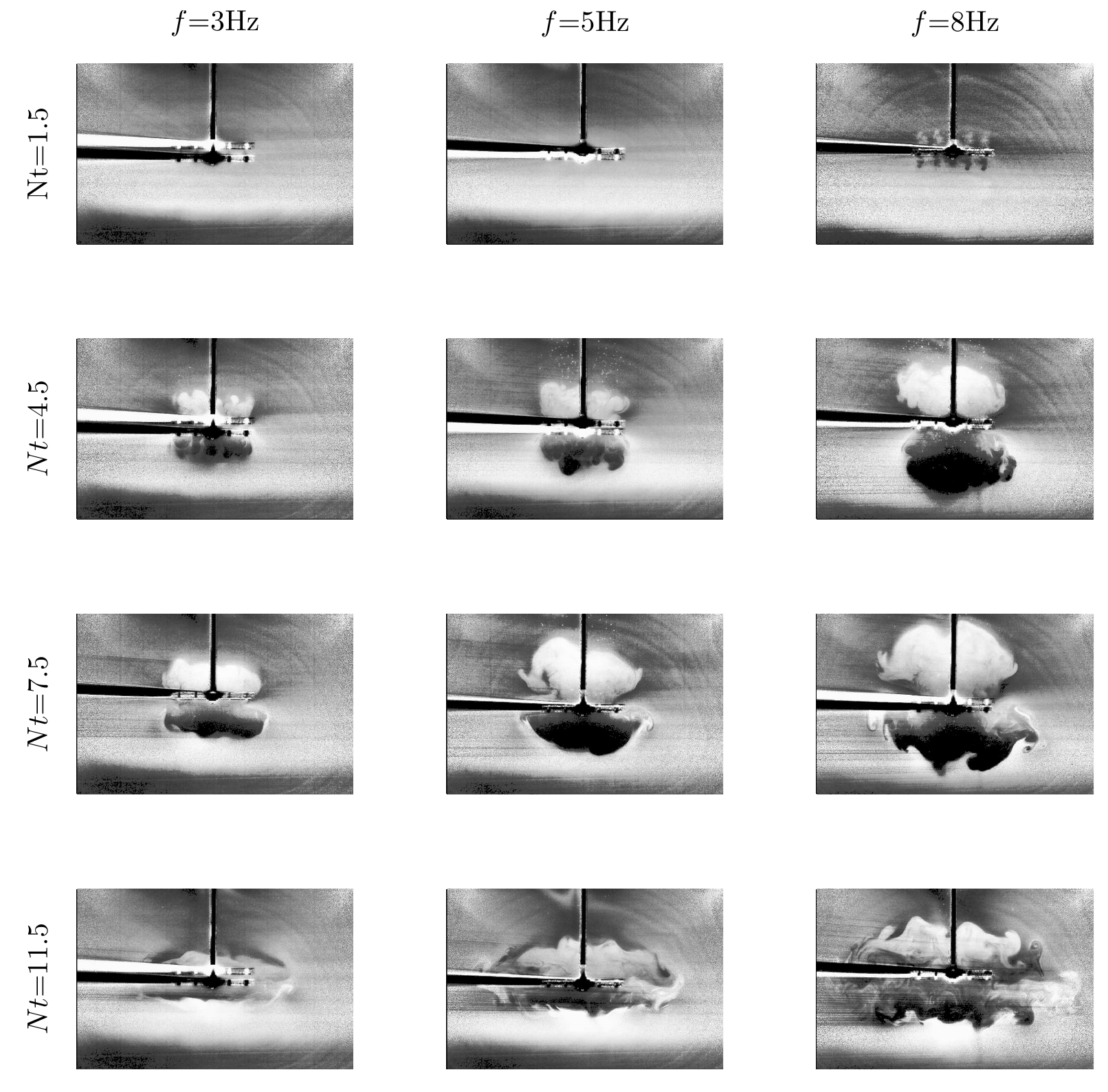}
\caption{PLIF images showing the development of the turbulent patch at different time instants $Nt = 1.5,4.5,7.5,11.5$ for the grid 
frequencies $f = 3, 5, 8$ Hz.}
\label{fig:density} 
\end{figure}

The patch evolution in time is shown from PLIF images in \figLabel\ref{fig:density} for different time instants normalised by the stratification frequency, $N t$, and for the representative forcing frequencies $f = 3, 5,$ and  $8 \unit{Hz}$ ($Nt = 0$ is the time when the grid starts from rest). For the sake of clarity, these are not the raw PLIF images. The first image (corresponding to the initially undisturbed concentration field of a linear stable stratified fluid at the dimensionless time $N t$=0) was subtracted from raw images. The resulting images are also processed with the noise low pass filter, and adjusted for brightness and contrast enhancement.

As depicted in \figLabel\ref{fig:density}, at early times, individual vortices emanating from the grid appear (particularly visible for the $f=8$ Hz) that rapidly merge and form a uniformly mixed fluid which moves away from the grid. As time progresses, counter-rotating structures are formed, pulling fluid inward from the sides at the level of the grid, and ejecting fluid outwards away from the grid due to baroclinic torque. The patch attains a maximal vertical size at $Nt\approx 5.5$ for all frequencies, with the patch size growing with increasing the forcing frequency which is consistent with earlier observations~\cite{vandeWatering1966, Merritt1974, Lin1979,DeSilva1998}. 

The turbulent/non-turbulent interface exhibits relatively small undulations.  The undulations become much more prominent upon the patch collapse which occurs after the grid is switched off at $Nt=8$. At this late stage, the formation of an intrusive gravity current is observed. Note that the wave-like features that seem to be radiating away from the turbulent region are artefacts associated with the imaging and not related to internal gravity waves. Indeed, the artefacts are visible in the thoroughly mixed and uniform density calibration images as well and are most likely caused by an optical filter, which was slightly smaller than the camera lens.

Raw images like those shown in \figLabel\ref{fig:density} are converted into density fields $\rho(x,z,t)$ through the PLIF calibration procedure (described in Section~\ref{sec:plif}) and transformed into the buoyancy fields according to the following relation:
\begin{equation}
\label{eq:bdef}
b = g \frac{\rho_0 - \rho}{\rho_0}
\end{equation}
Here it should be noted that density variations are small such that $\rho_0 - \rho$ can be interpreted as a perturbation buoyancy consistent with the Boussinesq approximation.

\begin{figure}
\centering
\includegraphics[width=\textwidth]{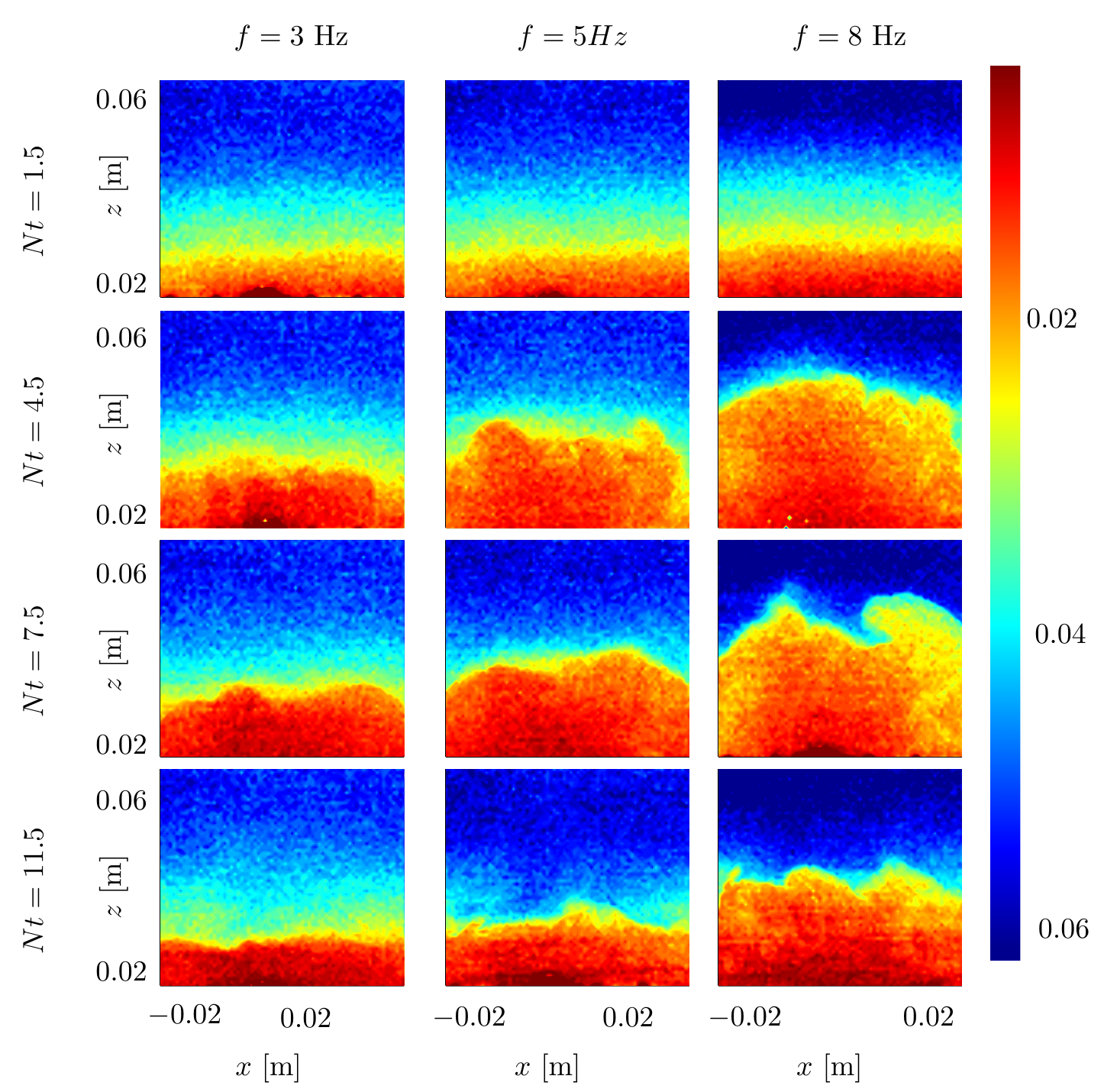}
\caption{Time evolution of the buoyancy $b(z,t)$ [m/s$^2$] for forcing frequencies of 3, 5, and 8 Hz. Note the positive direction of $z$ upon an invariant transformation is pointing upwards.}
\label{fig:buoyancymap}
\end{figure}

Figure~\ref{fig:buoyancymap} shows the zoom in view on buoyancy $b(x,z,t)$ in the region of interest (indicated in \figLabel\ref{fig:comb}) for the experiment performed at $f=5$ Hz. As shown in Figs.~\ref{fig:density}-\ref{fig:buoyancymap}, the patch is relatively well-mixed. The TNTI seems to sharpen in time, and undulations of the interface are small due to the relatively strong stratification (large values of the bulk Richardson number $Ri$). 

\subsection{Spatially averaged description of the patch}\label{sec:spatial_average}

In this section we present spatially (horizontally or isopycnally) averaged statistics of the flow in the quasi-homogeneous region shown in \figLabel\ref{fig:density}. The spatial averages are denoted by angular brackets $\avx{\cdot}$ (without a subscript) and depend on $z$ and $t$ only. Note that the results shown from \figLabel \ref{fig:buoyancymap} onwards will be presented as if the measurements were made above the grid. This is permitted as the governing equations are invariant under the transformation $\{z \rightarrow -z$, $w \rightarrow -w$, $b \rightarrow -b\}$. 

\begin{figure}
\centering
\includegraphics[width=\textwidth]{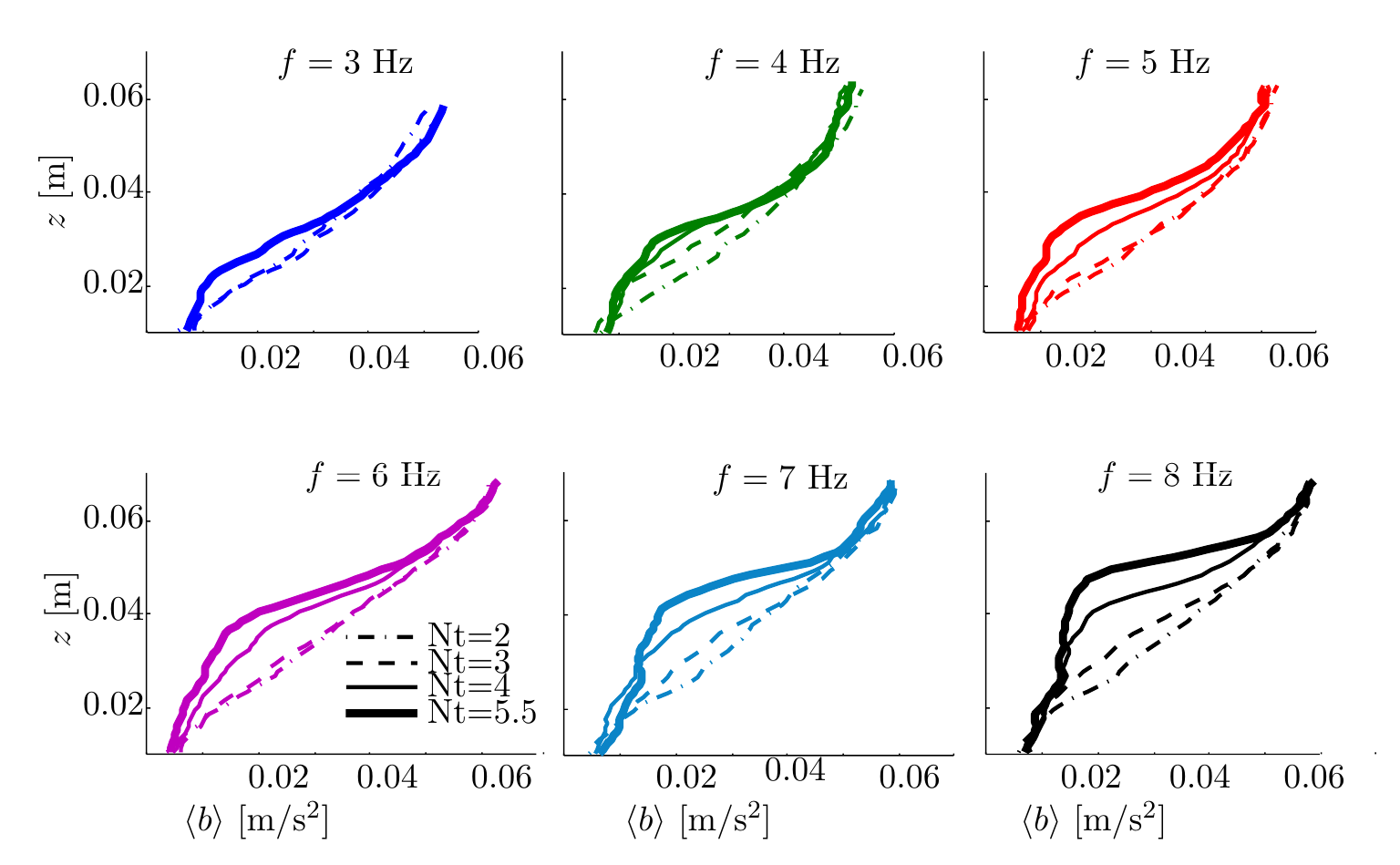}
\caption{Buoyancy profiles $\avx{b(z)}$ at different time instants, $Nt = 2,3,4$, and 5.5. Note the three slopes: the well-mixed patch, across the interface and the background buoyancy. \label{fig:b(z)}}
\end{figure}

The spatially averaged buoyancy profiles $\avx{b}$ as a function of time and $z$ are depicted in \figLabel\ref{fig:b(z)}, using the data of all frequencies for several representative time instants. The initial condition is the background stratification given by $\avx{b}=N^2 z$ where $N^2=1$ s$^{-1}$. 

The buoyancy profiles $\avx{b}$ in \figLabel\ref{fig:b(z)} can be decomposed into two three distinct regions according to the slopes: \emph{i}) inside the patch where the turbulent mixing decreases the stratification; \emph{ii}) the TNTI that develops starting from $N t=2$ (dash-dot line) in which the local values for $N^2$ are much larger than the background stratification, and \emph{iii}) the background ambient with the slope close to $N^2 \approx 1$ s$^{-1}$.   
The TNTI can be seen to move away from the grid until $N t=5.5$ (thick line), indicative of the turbulent entrainment that is taking place. In this period the TNTI is observed to sharpen, as we demonstrate below. Note that the values of $N^2$ inferred from the spatially averaged statistics likely underestimate the local values, due to the large-scale deformation of the flow near the interface, visible in \figLabel\ref{fig:density}. Local analysis across the interface is discussed in Section~\ref{sec:local} and it provides better estimates of the real TNTI interface thickness. 

\begin{figure}
\centering
\centering
\includegraphics[width=.7\textwidth]{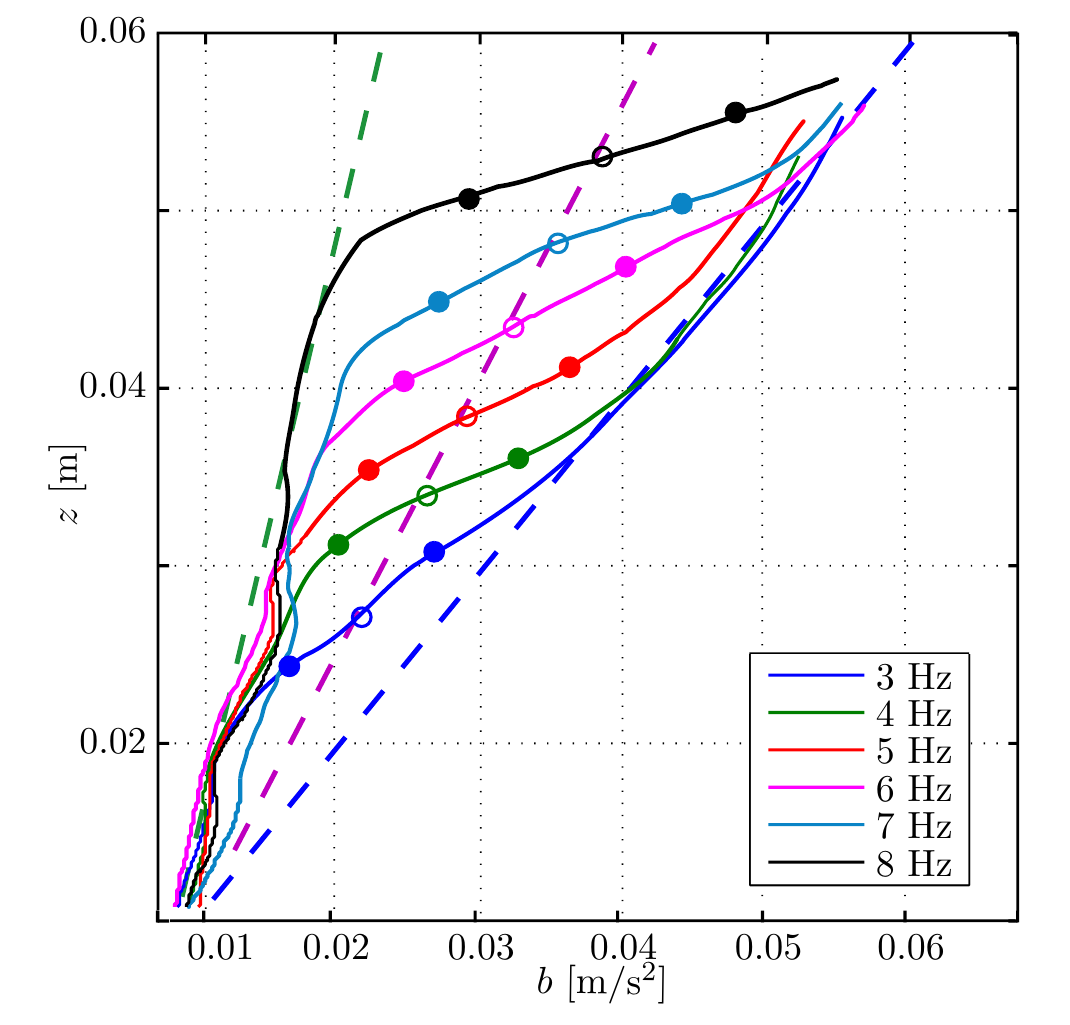}
\caption{The average buoyancy profiles at $Nt=5.5$ when the patch size is maximal, plotted together with the points identified using the interface detection technique. The dashed slopes are 0.3, 0.65 and 1, from bottom to top, respectively.  \label{fig:interface_identification} }
\end{figure}

In order to define the patch size, $\delta$ and the interface thickness $\Delta$, we use the notion that the stratification inside the patch is characterized by $N^2\approx 0.3$ s$^{-1}$ for all forcing frequencies, and that the background stratification is defined by $N^2 = 1$ s$^{-1}$ (\figLabel\ref{fig:interface_identification}). The slope across the interface is approximately 0.65 and it is shown by a middle dashed line in \figLabel\ref{fig:interface_identification}.  The intersection of each averaged profile $\avx{b}$ with this line (empty circles) is defined here as the mid-point of the interface and defines the patch size $\delta(f,t)$. Following Craske et al.\cite{Craske2015}, we decide to use 25\% and 75\% of the value at the intersection to mark the upper and the lower limits of the layer (full symbols) that we call hereinafter the ``interface layer''. The distance between these two limits on the profile serves as an estimate of the interface layer thickness, $\Delta(f,t)$. This approach to measure the position and the thickness of the interfacial layer is relatively robust and much less sensitive to experimental noise than other interface detection techniques (for instance using fixed thresholds or marking the maximum buoyancy gradient $d\avx{b}/dz$). 

\begin{figure}
\centering
\includegraphics[width=.7\textwidth]{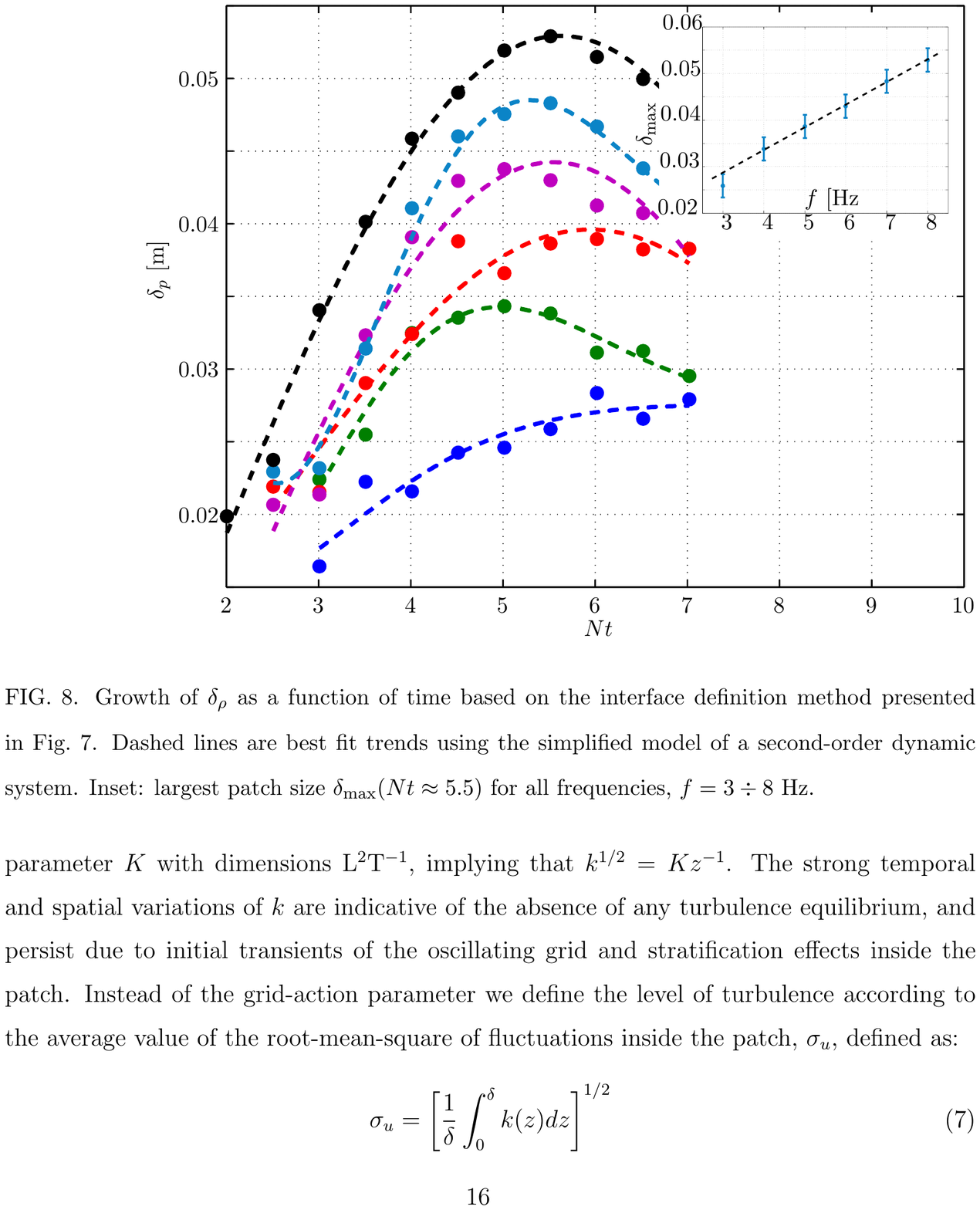}
\caption{\label{fig:layergrowth} Growth of $\delta_\rho$ as a function of time based on the interface definition method presented in \figLabel\ref{fig:interface_identification}. Dashed lines are best fit trends using the simplified model of a second-order dynamic system. Inset: largest patch size $\delta_{\mathrm{max}}(Nt \approx 5.5)$ for all frequencies, $f = 3\div 8$ Hz.}
\end{figure}

Figure~\ref{fig:layergrowth} presents the results of the patch growth in terms of $\delta(t)$, estimated using the aforementioned method (see \figLabel\ref{fig:interface_identification}) applied to the spatially averaged buoyancy profiles, shown in \figLabel\ref{fig:b(z)}. The results are shown versus dimensionless time $N t$, for each frequency. The patch size grows rapidly until $N t \approx 4$ after which the growth rate decreases drastically, to half of its original value and than the patch grows again until $N t \approx 5.5$ and reaches its maximum size. After that, the patch size declines until $N t \approx  8$, the time when the grid stops. Note that the patch collapses before we stop the motor and the evolution cycle (transient, initial growth and the maximum size) occur at the same time independent of the grid frequency. The behaviour of the patch size resembles a step response of a second-order dynamical system (a pendulum-like motion) with an overshoot type of under-damped system response for higher frequencies and over-damped behaviour for the lower frequencies. The curves through the experimental points are best fits of such a solution, emphasizing the similarity of the behaviour at different frequencies. 

The inset of \figLabel\ref{fig:layergrowth} shows the relation between the maximum patch size, $\delta_{\mathrm{max}}(Nt = 5.5)$ and the forcing frequency, $f$. A linear trend is discernible, which is consistent with earlier observation by Van de Watering~\cite{vandeWatering1966} and Fernando~\cite{Fernando1988}. Also it is apparent that the patch largest size value is directly proportional to the forcing frequency.

\subsection{Spatially averaged and turbulent flow statistics}

\begin{figure}
\centering
\centering
\includegraphics[width=\textwidth]{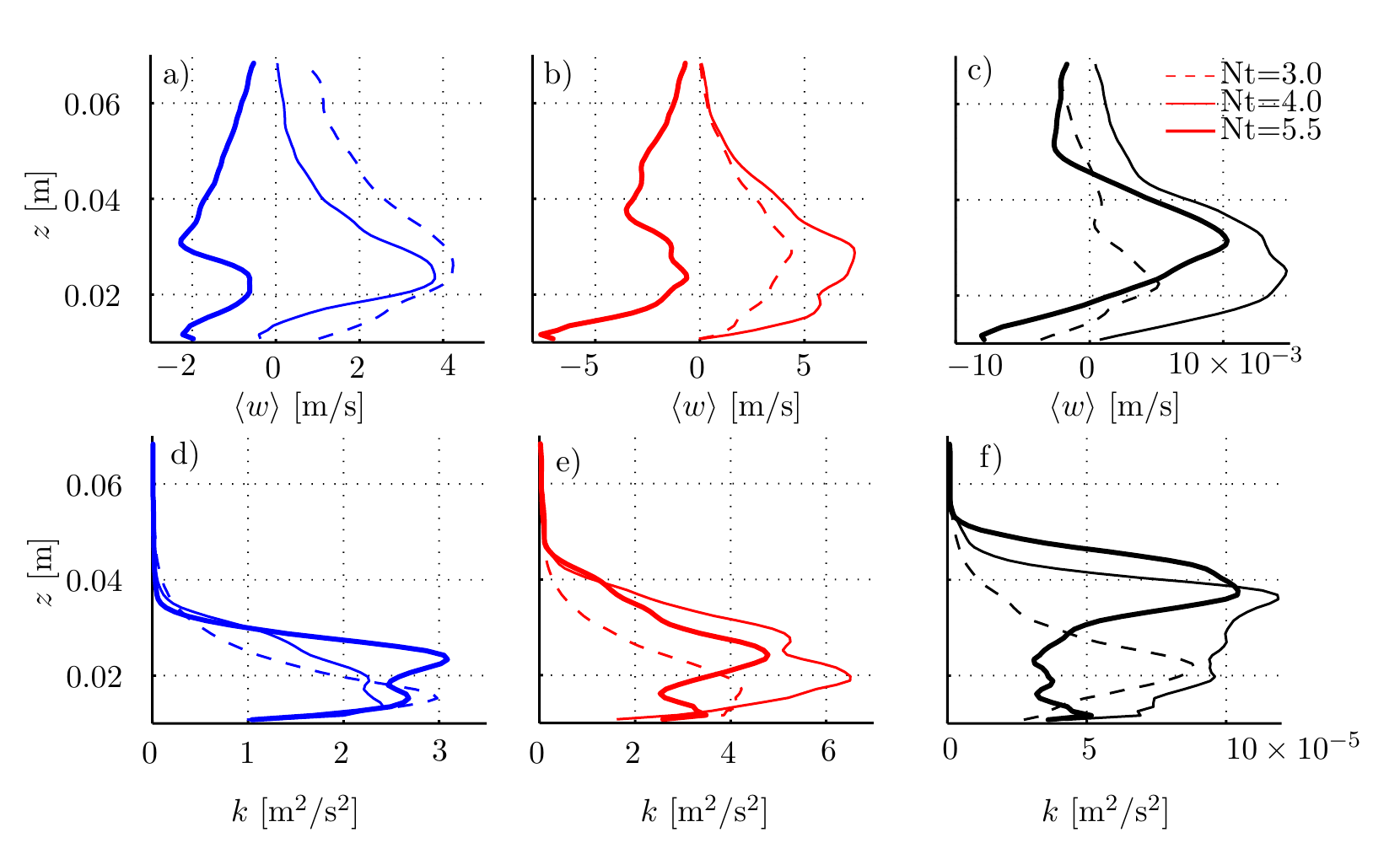}
\caption{\label{fig:transients} Time-evolution of spatially averaged profiles for all frequencies for $Nt = 3, 4, 5.5$ (dashed, thin solid and the thick solid lines, respectively) of (a-c) spatially averaged vertical velocity $\avx{w}$ and (d-f) turbulent kinetic energy $k = \frac{1}{2} \avx{\fl{u}^2 + \fl{w}^2}$. Color coding is identical to \figLabel\ref{fig:interface_identification}, (left) 3 Hz (center) 5 Hz and (right) 8 Hz, respectively.} 
\end{figure}

The spatially averaged vertical velocity $\avx{w}$ is shown in \figLabel\ref{fig:transients} (a-c) at three instants in time for the forcing frequencies 3, 5 and 8 Hz. Positive vertical velocity for the first two time instants is indicative of a mean flow away from the grid which is consistent with the observed counter-rotating rolls at the edges of the grid discernible in \figLabel\ref{fig:density}. At $Nt = 5.5$, the mean velocity has changed sign for both the $f=3$ and $5$ Hz cases, and partly for $f=8$ Hz.The change of sign is due to the presence of well-mixed fluid inside the patch which creates hydrostatic pressure values larger than the ambient which in turn will cause the formation of an intrusive gravity current.
In addition to the mean flow we quantify the turbulent kinetic energy (TKE). The spatially averaged TKE is defined here as $k = \frac{1}{2} \avx{u'^2 + w'^2}$ where $\fl{u} = u - \avx{u}$ and $\fl{v} = v - \avx{v}$, and it is shown in \figLabel\ref{fig:transients} (d-f). The TNTI is emphasised in these figures by the sharp drop-off of TKE near the interface. The slope of this decrease of TKE across the interface layer is increasing with the intensity of the patch forcing which is in our case with frequency.  

\begin{figure}
\centering
\includegraphics[width=.6\textwidth]{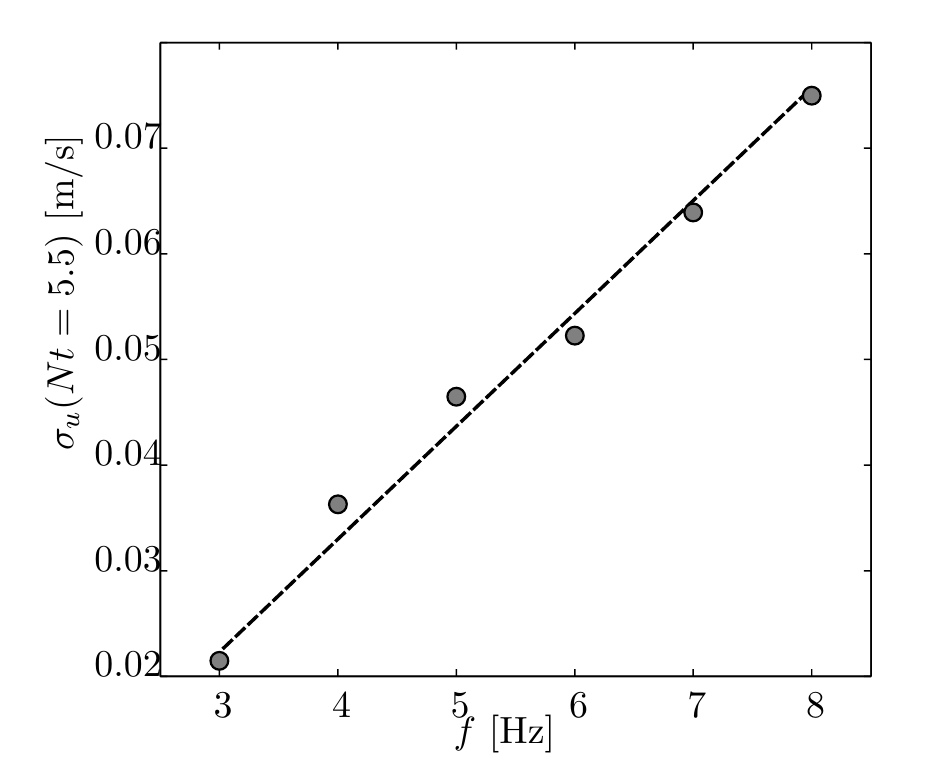}
\caption{\label{fig:sigmau_vs_freq} Measure of turbulent velocity fluctuations inside the patch, $\sigma_u$ versus the frequency of the grid, $f$, estimated according to Eq.~\ref{eq:sigma_u}. Dashed line is a linear best fit trend.}
\end{figure}

Figure~\ref{fig:transients}(d-f) show that the shape of the turbulence profiles varies strongly with time.
This means that we cannot use the commonly used "grid action" description \cite{Long1978, DeSilva1992, DeSilva1998}, which assumes that the turbulence generated by the grid can be described by a grid action parameter $K$ with dimensions L$^2$T$^{-1}$, implying that $k^{1/2} = K z^{-1}$. The strong temporal and spatial variations of $k$ are indicative of the absence of any turbulence equilibrium, and persist due to initial transients of the oscillating grid and stratification effects inside the patch. Instead of the grid-action parameter we define the level of turbulence according to the average value of the root-mean-square of fluctuations inside the patch, $\sigma_u$, defined as: 
\begin{equation}\label{eq:sigma_u}
\sigma_u = \left[\frac{1}{\delta} \int_{0}^\delta k (z) dz \right]^{1/2} 
\end{equation}
\noindent Dependence of $\sigma_u$ on the forcing frequency $f$ at $Nt=5.5$ is shown in \figLabel\ref{fig:sigmau_vs_freq}. The result shows clearly that the level of velocity fluctuations $\sigma_u$ is linearly dependent on the forcing frequency, consistent with previous studies \cite{DeSilva1992, DeSilva1998} that used oscillating grids.  Note that the linear relation between $\sigma_u$ on frequency is also consistent with the observed linear dependence of $\delta_{\mathrm{max}}$ and $f$. Indeed, based on dimensional arguments, we expect that $\delta \sim \sigma_u/N$, which upon substituting $\sigma_u \sim f$ results in $\delta \sim f / N$ (see \figLabel\ref{fig:layergrowth}).

\subsection{Local analysis of entrainment across the interface}\label{sec:local}

\begin{figure}
\includegraphics[width=\textwidth]{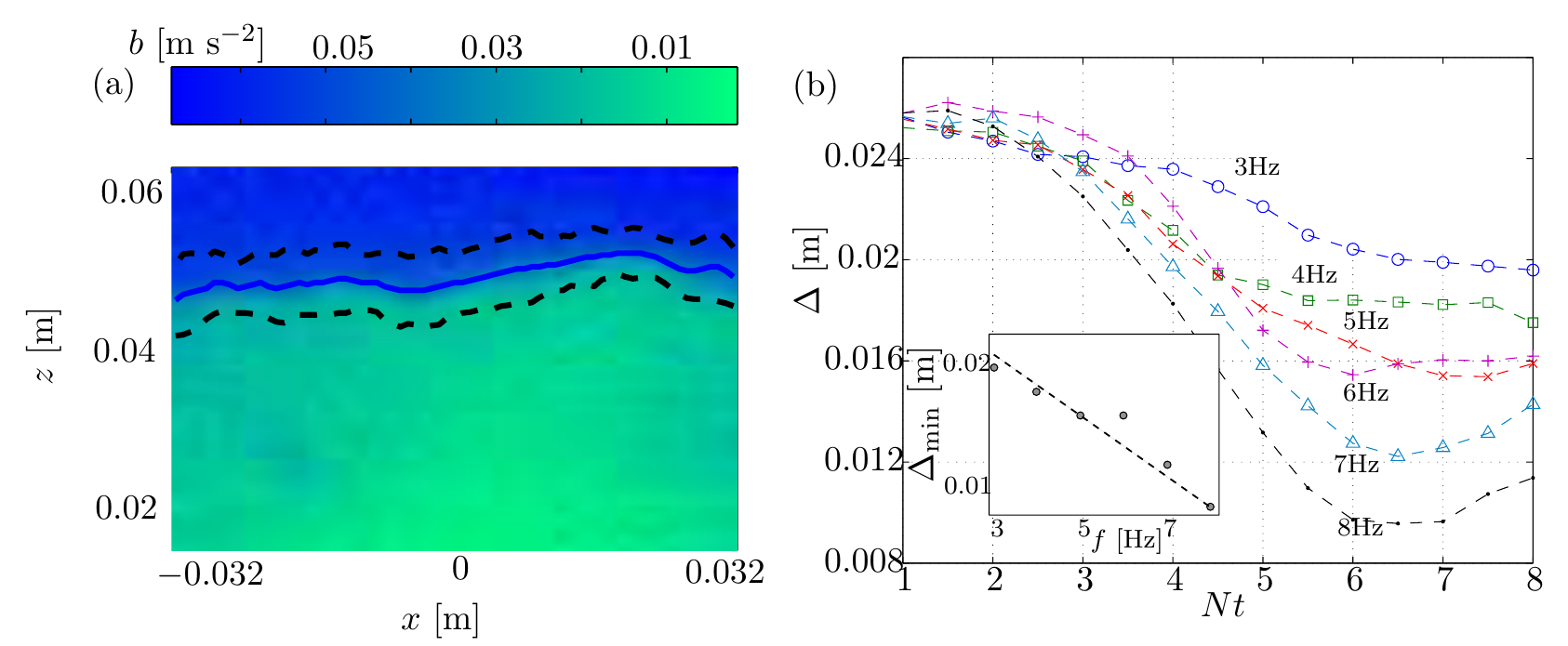}
\caption{(a) example buoyancy map for 8Hz, at time $N t = 5.5$ with the marked interface position and thickness of the interface, $\Delta(x)$ defined according to the $25\div 75\%$. (b) thickness of the interface $\langle \Delta \rangle (f,t)$ for different frequencies versus dimensionless time. Inset: minimal thickness vs frequency. the fit is $D_{\mathrm{min}}$. Color coding is identical to \figLabel\ref{fig:interface_identification}. \label{fig:interface_thickness}}
\end{figure}

Using the interface detection method (presented in \figLabel\ref{fig:interface_identification}) we also estimate the position of the interface, $\delta(x,t)$ as well as the local thickness of the interface, $\Delta(x,t)$. In \figLabel\ref{fig:interface_thickness}(a) we demonstrate the buoyancy field (as a color map) on which we mark the location of the interface (solid curve), and the corresponding 25\% and 75\% of the interface value (dashed lines). The distance between the two dashed lines is the local thickness of the interface, $\Delta(x,t)$.  

The values of the interface thickness $\Delta(x,t)$, measured at different $x$ locations and different time instants, are then averaged along $x$ and plotted in \figLabel\ref{fig:interface_thickness}(b). It appears that the thickness of the interface is decreasing with increasing frequency and the interface sharpens with time, at least until the $Nt \approx 5.5$ that corresponds to the critical moment when the growth stops and the patch collapses. It is noteworthy that at this moment, the size of the patch is maximal and the thickness is minimal. Since the patch is well-mixed, it means that the patch reaches the largest gradient of buoyancy at this time instant of $Nt \approx 5.5$. Plotting the minimal value of $\avx{\Delta}$ versus frequency (inset \figLabel\ref{fig:interface_thickness}(b)) also results in an approximately linear dependence of $\Delta_{\textnormal{min}}$ on $f$. 

\begin{figure}
\centering
\includegraphics[width=\textwidth]{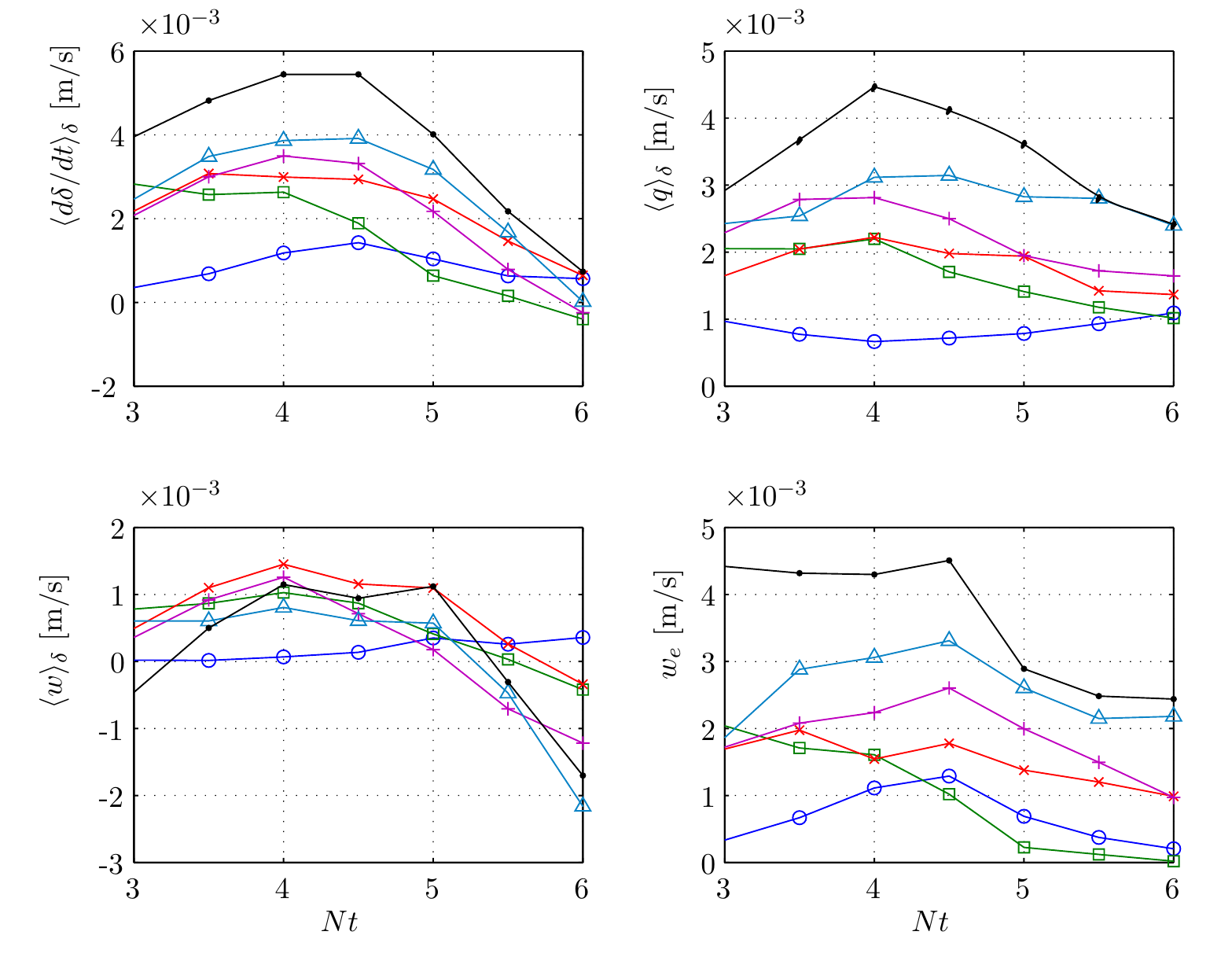}
\caption{Locally measured quantities: (a) rate of growth of a patch, $\avx{d\delta/dt}_{\delta}$, (b) square root of the available kinetic energy at the interface, $\avx{k}_{\delta}$ (c) average vertical velocity $\avx{w}_{\delta}$ and (d) entrainment velocity, $w_e$, all in units of [m/s].  \label{fig:figE}}
\end{figure}

We show in the following how the turbulent entrainment is associated with the evolution of the patch vertical size with time. In the previous section it was shown that there is a substantial component of the spatially averaged vertical flow $\avx{w}$ inside the patch, implying that the propagation of the interface due to turbulent mixing is: 
\begin{equation}
w_e(t) = \left \langle \frac{d \delta}{\d t}\right \rangle_{\delta} - \avx{w}_{\delta}
\end{equation}
\noindent Subscript $\delta$ implies averaging along the interface defined by $\delta(x,t)$, according to  $\avx{\chi}_\delta \equiv 1/L \int_{-L/2}^{L/2} \chi(x, \delta(x), t) \d x$ where $\chi$ is the averaged variable.
The quantities  $\avx{d \delta / d t}_\delta$, $\avx{w}_\delta$ and the entrainment velocity $w_e$ are plotted in \figLabel\ref{fig:figE}a, c, d for all the experiments. 

It seen that $w_e$ is largely positive as is expected: the turbulence mixing is pushing the interface outwards as long as the grid is oscillating. In addition, as $w_e$ remains positive for all time, the patch collapse must be caused by mean flow effects.

For the sake of local balance, we use a local turbulent velocity scale $\avx{q}_\delta$, where $q = k^{1/2}$ is the square root of the TKE. The local turbulent velocity scale is shown in \figLabel\ref{fig:figE}b and its comparison with the entrainment velocity $w_e$ (\figLabel\ref{fig:figE}d) emphasize some degree of correlation between these key parameters.

\begin{figure}
\centering
\includegraphics[width=\textwidth]{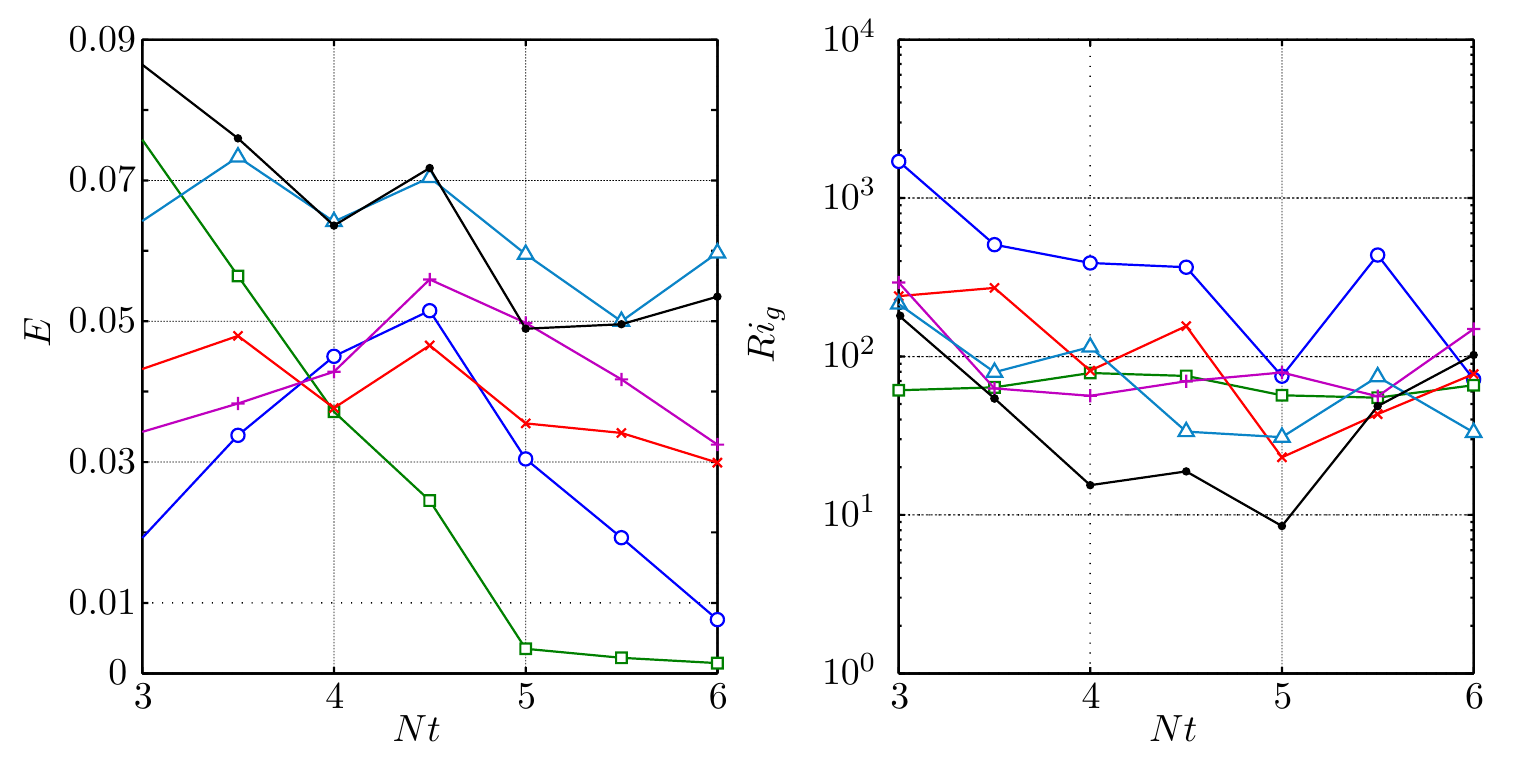}
\caption{\label{fig:E_Ri} (a) Entrainment rate, $E$ as a function of time and (b) Local gradient richardson number $Ri_g = \avx{\d b/\d z}/\avx{\d q/\d z}^2$ where $q = \sqrt{k}$. Color coding is identical to \figLabel\ref{fig:interface_identification}.}
\end{figure}

Using the local parameters from \figLabel\ref{fig:figE}, it is possible to extract a local entrainment coefficient $E$, defined as:
\begin{equation}
E = \frac{w_e}{q}
\end{equation}
\noindent which is plotted in \figLabel\ref{fig:E_Ri}a as a function of time for all frequencies. It is observed that $E$ increases as the forcing frequency is increased. Furthermore, the value of $E$ seems to drop as time progresses. These large variations are to be expected due to the non-equilibrium character of turbulence and of particular relevance here is that the value of $E$ is positive definite and constrained to the interval $E \le 0.1$.

The entrainment coefficient is typically assumed to be dependent on the Richardson number $Ri$ in a form of the entrainment law, $E=E(Ri)$ \cite{Fernando1991}. The definition of $Ri$ vary from one flow type to another\cite{Fernando1991}. The initial growth of a localised patch under consideration does not support such an entrainment law, which is not surprising as this problem features large non-equilibrium effects both in the mean flow and the turbulence. Nevertheless, it is instructive to calculate a local gradient Richardson number defined as:
\begin{equation}
Ri_g = \frac{\avx{\d b/\d z}_\delta}{\avx{\d q/\d z}_\delta^2}
\end{equation}
From the data shown in \figLabel\ref{fig:E_Ri}a it is evident that the values of $Ri_g$ at the interface are of the order of $\mathcal{O}(10^2)$, implying that the turbulence is strongly suppressed in the vertical direction. Among the different runs, the values of $Ri_g$ for the $f=3$ Hz experiment are the highest. This may be counter-intuitive, as one would expect that lower forcing would produce interfaces that are less sharp; however, in this case, it is $(\d q / \d z)^2$ which becomes small at the TNTI much quicker than $(\d b / \d z)$. It is difficult to conclude about $Ri_g$ dependence on frequency, although it seems that the value of $Ri_g$ stays fairly constant as a function of time and does not vary strongly with $f$.


\section{Concluding remarks}\label{sec:conclusions}

We have performed simultaneous PIV and PLIF experiments in a localized turbulent patch, freely developing far from the walls in a stably stratified environment.
Consistent with previous studies we observe that the patch vertical size, $\delta(t)$, initially grows rapidly, then reaches a maximum, followed by a collapse and subsequent equilibrium through the formation of a gravity current. The time it takes to reach the maximum size is $Nt \approx 5.5$, independent of frequency, whilst the patch growth rate changes drastically at $Nt \approx 4$.

The focus of this paper is on the early transients of the patch that occur before the quasi-steady state occurs, i.e. $Nt < 8$. The measurements reveal significant large-scale flow features in the form to counter-rotating vortices on the lateral patch edges after the grid was switched on, which are both visible directly in the buoyancy field (\figLabel \ref{fig:density}) and indirectly in the form of a mean velocity away from the grid in the central area of the patch (\figLabel \ref{fig:figE}c).
The turbulence intensity inside the turbulent patch is large, exceeding the mean flow for all frequencies.
Near the turbulent/non-turbulent interface, turbulent levels dropped off rapidly, whilst the mean flow components extend beyond the turbulent region.

The data reveal strong non-equilibrium behaviour both for the mean flow and the turbulence. This implies that the concept of grid action, commonly used to characterise the turbulence intensity for oscillating grids, is not valid for the early dynamics analysis of a localised patch. Other factors that may be contributing are the stratification inside the patch and the fact that the patch is typically not large enough to reach the several mesh sizes required to establish a behaviour consistent with the grid action concept.  

A good indicator for the turbulence level inside the patch, $\sigma_u$ is defined in this work based on the spatial average inside the well-mixed region. It is found that $\sigma_u$ at $Nt \approx 5.5$ is directly proportional to the grid forcing frequency, $f$, in agreement with previous studies. 

The local analysis at the interface is carried out, revealing to what extent the TNTI propagates by advection due to the mean flow and to what extent by turbulence and turbulent entrainment across the interface. It is observed that advection by the mean flow is the process responsible for the sign reversal in $\d \delta / \d t$ after $N\,t \approx 5.5$ and the following collapse. Importantly, the entrainment velocity remains positive as turbulence consistently causes outward propagation of the interface due to turbulent mixing. The values of the entrainment rate $E$ are found to be constrained to $E \le 0.1$. The gradient Richardson number $\Ri_g$ at the interface is rather high, $\mathcal{O}(10^2)$, implying that suppression of the vertical turbulence components are  significant due to very sharp interface and extremely strong buoyancy gradient across the interface.

\section*{Acknowledgments}
The authors acknowledge support by the Bi-National U.S. Israel Science Foundation Grant 2008051 and Israel Science Foundation Grant no. 945/15. Z.J. Taylor is grateful to the Azrieli Foundation for the award of an Azrieli Fellowship. 

\bibliographystyle{apalike}
\bibliography{patch}

\begin{thebibliography}{}

\bibitem[Alahyari and Longmire, 1994]{Alahyari1994}
Alahyari, A. and Longmire, E. (1994).
\newblock Particle image velocimetry in a variable density flow: application to
  a dynamically evolving microburst.
\newblock {\em {Exp. Fluids}}, 17:434--440.

\bibitem[Alford and Gregg, 2001]{AlfordGregg2001}
Alford, M.~H. and Gregg, M.~C. (2001).
\newblock {Near-inertial mixing: Modulation of shear strain and micro-structure
  at low latitude}.
\newblock {\em J. Geophys. Res.}, 106:16947--16968.

\bibitem[Alford and Pinkel, 2000]{AlfordPinkel2000}
Alford, M.~H. and Pinkel, R. (2000).
\newblock {Observations of overturning in the thermocline: The context of ocean
  mixing}.
\newblock {\em J. Phys. Oceanogr.}, 5:805--832.

\bibitem[Arneborg, 2002]{Arneborg2002}
Arneborg, L. (2002).
\newblock Mixing efficiencies in patchy turbulence.
\newblock {\em J. Phys. Oceanogr.}, 32:1496--1506.

\bibitem[Atsavapranee and Gharib, 1997]{Atsavapranee1997}
Atsavapranee, P. and Gharib, M. (1997).
\newblock {Structures in stratified plane mixing layers and the effects of
  cross-shear}.
\newblock {\em J. Fluid Mech.}, 342:53--86.

\bibitem[Broutman, 1986]{Broutman1986}
Broutman, D. (1986).
\newblock On internal wave caustics.
\newblock {\em J. Phys. Oceanogr.}, 16:1625--1635.

\bibitem[Craske et~al., 2015]{Craske2015}
Craske, J., Debugne, A. L.~R., and {van Reeuwijk}, M. (2015).
\newblock {Shear-flow dispersion in turbulent jets}.
\newblock {\em J. Fluid Mech.}, 781:28--51.

\bibitem[Crimaldi, 2008]{Crimaldi2008}
Crimaldi, J.~P. (2008).
\newblock {Planar laser induced fluorescence in aqueous flows}.
\newblock {\em Exp. Fluids}, 44:851--863.

\bibitem[Daviero et~al., 2001]{Daviero2001}
Daviero, G.~J., Roberts, W., and Maile, K. (2001).
\newblock Refractive index matching in large-scale stratified experiments.
\newblock {\em Exp. Fluids}, 31:119--126.

\bibitem[{De Silva} and Fernando, 1992]{DeSilva1992}
{De Silva}, I. P.~D. and Fernando, H. J.~S. (1992).
\newblock Some aspects of mixing in a stratified turbulent patch.
\newblock {\em J. Fluid Mech.}, 240:601--625.

\bibitem[{De Silva} and Fernando, 1998]{DeSilva1998}
{De Silva}, I. P.~D. and Fernando, H. J.~S. (1998).
\newblock Experiments on collapsing turbulent regions in stratified fluids.
\newblock {\em J. Fluid Mech.}, 358:29--60.

\bibitem[Economidou and Hunt, 2009]{Economidou2009}
Economidou, M. and Hunt, G.~R. (2009).
\newblock Density stratified environments: the double-tank method.
\newblock {\em Exp. Fluids}, 46:453--466.

\bibitem[Fernando, 1988]{Fernando1988}
Fernando, H. J.~S. (1988).
\newblock The growth of a turbulent patch in a stratified fluid.
\newblock {\em J. Fluid Mech.}, 190:55--70.

\bibitem[Fernando, 1991]{Fernando1991}
Fernando, H. J.~S. (1991).
\newblock Turbulent mixing in stratified fluids.
\newblock {\em Ann. Rev. Fluid Mech.}, 23:455--493.

\bibitem[Ferrier et~al., 1993]{Ferrier1993}
Ferrier, A.~J., Funk, D.~R., and Roberts, P. J.~W. (1993).
\newblock Application of optical techniques to the study of plumes in
  stratified flows.
\newblock {\em Dyn. Atmos. Oceans}, 20:155--183.

\bibitem[Ferron et~al., 1998]{Polzin1998}
Ferron, B., Herle, M., Speer, K., Gargett, A., and Polzin, K. (1998).
\newblock {Mixing in the Romanche fracture zone}.
\newblock {\em J. Phys. Oceanogr.}, 28:1929--1945.

\bibitem[Garrett and Laurent, 2002]{GarrettLaurent2002}
Garrett, C. and Laurent, L. (2002).
\newblock Aspects of deep ocean mixing.
\newblock {\em J. Oceanogr.}, 58:11--24.

\bibitem[Garrett and Munk, 1972]{Garrett1972}
Garrett, C. and Munk, M. (1972).
\newblock Oceanic mixing by breaking internal waves.
\newblock {\em Deep-Sea Res.}, 19:823--832.

\bibitem[Garrett and Munk, 1979]{Garrett1979}
Garrett, C. and Munk, W. (1979).
\newblock Internal waves in the ocean.
\newblock {\em {Ann. Rev. Fluid Mech.}}, 11:339--369.

\bibitem[Klymak and Moum, 2007]{KlymakMoum2007}
Klymak, J.~M. and Moum, J.~N. (2007).
\newblock Oceanic isopycnal slope spectra: Part ii-turbulence.
\newblock {\em J. Phys. Oceanogr.}, 37:1232--1245.

\bibitem[Li and Yamazaki, 2001]{LiYamazaki2001}
Li, H. and Yamazaki, H. (2001).
\newblock {Observations of a Kelvin-Helmholtz billow in the ocean}.
\newblock {\em J. Oceanogr.}, 57:709--721.

\bibitem[Lin and Pao, 1979]{Lin1979}
Lin, J.~T. and Pao, Y.~H. (1979).
\newblock Wakes in stratified fluids.
\newblock {\em {Ann. Rev. Fluid Mech.}}, 11(1):317--338.

\bibitem[Long, 1978]{Long1978}
Long, R.~R. (1978).
\newblock Theory of turbulence in a homogeneous fluid induced by an oscillating
  grid.
\newblock {\em Phys. Fluids}, 21:1887--1888.

\bibitem[Marmorino, 1987]{Marmorino1987}
Marmorino, G. (1987).
\newblock {Observations of small-scale mixing processes in the season
  thermocline. Part II: Wave breaking.}
\newblock {\em J. Phys. Oceanogr.}, 17:1348--1355.

\bibitem[McDougall, 1979]{McDougall1979}
McDougall, T.~J. (1979).
\newblock On the elimination of refractive-index variations in turbulent
  density-stratified liquid flows.
\newblock {\em J. Fluid Mech.}, 93:83--96.

\bibitem[Merritt, 1974]{Merritt1974}
Merritt, G.~E. (1974).
\newblock {Wake growth and collapse in stratified flow}.
\newblock {\em AIAA J.}, 12(7):940--949.

\bibitem[Nash et~al., 2007]{Kelly2007}
Nash, J.~D., Alford, M.~H., Kunze, E., Martini, K., and Kelly, S. (2007).
\newblock {Hotspots of deep ocean mixing on the Oregon continental slope}.
\newblock {\em Geophys. Res. Lett.}, 34:01605.

\bibitem[Nasmyth, 1970]{Nasmyth1970}
Nasmyth, P.~W. (1970).
\newblock {\em Oceanic Turbulence}.
\newblock PhD thesis, University of British Columbia.

\bibitem[Odier et~al., 2009]{Odier2009}
Odier, P., Chen, J., Rivera, M.~K., and Ecke, R.~E. (2009).
\newblock {Fluid mixing in stratified gravity currents: the Prandtl mixing
  length}.
\newblock {\em Phys. Rev. Lett.}, 102:134504.

\bibitem[Peters et~al., 1995]{Sanford1995}
Peters, H., Gregg, M.~C., and Sanford, T.~B. (1995).
\newblock Detail and scaling of turbulent overturns in the pacific equatorial
  undercurrent.
\newblock {\em J. Geophys. Res.}, 100:18349--18368.

\bibitem[Sarathi et~al., 2011]{Sarathi2011}
Sarathi, P., Gurka, R., Kopp, G., and Sullivan, P. (2011).
\newblock A calibration scheme for quantitative concentration measurements
  using simultaneous {PIV} and {PLIF}.
\newblock {\em Exp. Fluids}, 52:247--259.

\bibitem[Smyth et~al., 2001]{Caldwell2001}
Smyth, W.~D., Moum, J.~N., and Caldwell, D.~R. (2001).
\newblock The efficiency of mixing in turbulent patches: inferences from direct
  simulations and microstructure observations.
\newblock {\em J. Phys. Oceanogr.}, 31(8):1969--1992.

\bibitem[Spedding, 1997]{Spedding1997}
Spedding, G.~R. (1997).
\newblock The evolution of initially turbulent bluff-body wakes at high
  internal froude number.
\newblock {\em J. Fluid Mech.}, 337:283--301.

\bibitem[Taylor et~al., 2010]{Taylor2010}
Taylor, Z.~J., Gurka, R., Kopp, G., and Liberzon, A. (2010).
\newblock {Long-duration time-resolved PIV to study unsteady aerodynamics}.
\newblock {\em IEEE Trans Instr. Meas.}, 59:3262--3269.

\bibitem[Troy and Koseff, 2005]{Troy2005}
Troy, C.~D. and Koseff, J.~R. (2005).
\newblock The instability and breaking of long internal waves.
\newblock {\em J. Fluid Mech.}, 543:107--136.

\bibitem[{van de Watering}, 1966]{vandeWatering1966}
{van de Watering}, W. P.~M. (1966).
\newblock The growth of a turbulent wake in a density-stratified fluid.
\newblock Technical Report 231-12, Hydronautics.

\bibitem[Wu, 1969]{Wu1969}
Wu, J. (1969).
\newblock Mixed region collapse with internal wave generation in a
  density-stratified medium.
\newblock {\em J. Fluid Mech.}, 35:531--544.

\end{thebibliography}

\end{document}